\documentclass[twocolumn,showpacs,preprintnumbers,amsmath,amssymb]{revtex4}

\usepackage{graphicx}
\usepackage{dcolumn}
\usepackage{bm}

\begin{document}
\title{Global spectral representations of black hole spacetimes in the complex plane}
\author{Maurice H.P.M. van Putten}
\affiliation{MIT-LIGO, 175 Albary Street, NW17-161, Cambridge, MA 02139}
\begin{abstract}
Binary black hole coalescence produces a finite burst of gravitational 
radiation which propagates towards quiescent infinity. These spacetimes
are analytic about infinity and contain a dimensionless coupling constant $M/s$, 
where $M$ denotes the total mass-energy and $s$ an imaginary distance. This
introduces globally convergent Fourier series on a complex radial coordinate,
allowing spectral representation of black hole spacetimes in all three dimensions.
We illustrate this representation theory on a Fourier-Legendre expansion of 
Boyer-Lindquist initial data and a scalar wave equation with signal recovery 
by Cauchy's integral formula.
\end{abstract}

\maketitle

\section{Introduction}

Rapid progress in sensitivity in the broadband gravitational-wave observatories
is creating new opportunities for searches for the bursts of gravitational-wave 
emissions produced by coalescing black holes. Detection strategies will benefit 
greatly from a priori understanding of their gravitational-wave emissions, which 
includes matched filtering techniques against a catalogue of signals precomputed 
by numerical relativity. This poses an interesting challenge of designing highly
efficient and stable computational algorithms.

Numerical relativity is particularly challenging in calculating
bursts and preceding chirps over many wave-periods in the presence of singularities 
associated with black hole spacetimes. Spectral methods provide an attractive 
approach. They are optimal in efficiency and accuracy, both in amplitude and phase, 
provided that the metric is smooth everywhere. Here, we focus on bursts
of radiation produced by black hole coalescence which propagate towards quiescent
infinity. By causality, these spacetimes preserve quiescence at infinity for all 
finite time.

The asymptotic structure of black hole spacetimes which are asymptotically flat 
and quiescent at infinity shares the same asymptotic properties as the Green's 
function of Minkowski spacetime. Consequently, these spacetimes carry a 
dimensionless coupling constant $M/s$, where $M$ denotes the total 
mass-energy of the space. The Green's function of the Laplacean,
\begin{eqnarray}
G(y^i,p^i)=\frac{1}{D(y^i,p^i)},
\end{eqnarray}
is a function of the distance $D$ between an observer at $y^i$ and a source at 
$p^i$. We can expand $G$ in Legendre polynomials $P_l$ using a spherical
coordinate system ($r,\theta,\phi)$, 
\begin{eqnarray}
G=\frac{1}{\sqrt{r^2+p^2-2pr\cos\theta}}=
p^{-1}\Sigma\left(\frac{p}{r}\right)^{2l+1} P_{2l}(\cos\theta),
\label{EQN_G2}
\end{eqnarray}
which shows that $G$ is analytic in each coordinate away from the singularity at 
$y^i=p^i$ with domain of convergence $r>p$. We introduce complex distances through 
a complex radial coordinate
\begin{eqnarray}
z=x+is~~(s>p).
\end{eqnarray}
The analytic continuation
\begin{eqnarray}
G=p^{-1}\Sigma \left(\frac{p}{z}\right)^{2l+1} P_{2l}(\cos\theta)
\label{EQN_GZ}
\end{eqnarray}
is globally convergent on $-\infty<x<\infty$ outside $|z|=p$. (This corresponding
branch cut for the square root is $(-\infty,0]$.) To be precise: 
for each $\theta$ (\ref{EQN_GZ}) defines a Taylor series in 1/z and, for each $z$, 
it defines a convergent expansion in Legendre functions. Thus, $p/s$ is a running
coupling constant and it must be less than one for convergence.
We shall refer to $M/s$ as the fixed coupling constant, noting
that $p/s$ is decreasing in binary coalescences.

Analyticity of the metric in Im(z)$\ge s$ defines an optimal distribution of 
points on $z=x+is$ for numerical implementation by the conformal transformation
\begin{eqnarray}
w=\frac{M}{z}
\end{eqnarray}
Here, $w$ describes a circle of radius $(2s)^{-1}$ about $w_0=(2is)^{-1}$,
whose interior corresponds to Im(z)$\ge s$. There exists a Taylor series 
in $w$ about $w_0$ which is convergent up to including the boundary
$C:~w=(2is)^{-1}\left(e^{i\phi}+1\right)$. The corresponding Fourier series 
in $\phi$ can be efficiently computed using the fast Fourier transform on a 
uniform distribution of points $\phi_k=2^{-m+1}\pi k$ ($k=1,2,\cdots, 2^m)$ 
on $C$. Their images on $z=x+is$ produce the non-uniform distribution 
\begin{eqnarray}
z_k=Ms\left[\tan(\phi_k/2)+i\right],
\label{EQN_ZK}
\end{eqnarray}
whose large-scale cut-off is set by the proximity of the $\{\phi_k\}$ to 
$\pm \pi$.

\begin{figure*}
\includegraphics[scale=0.9]{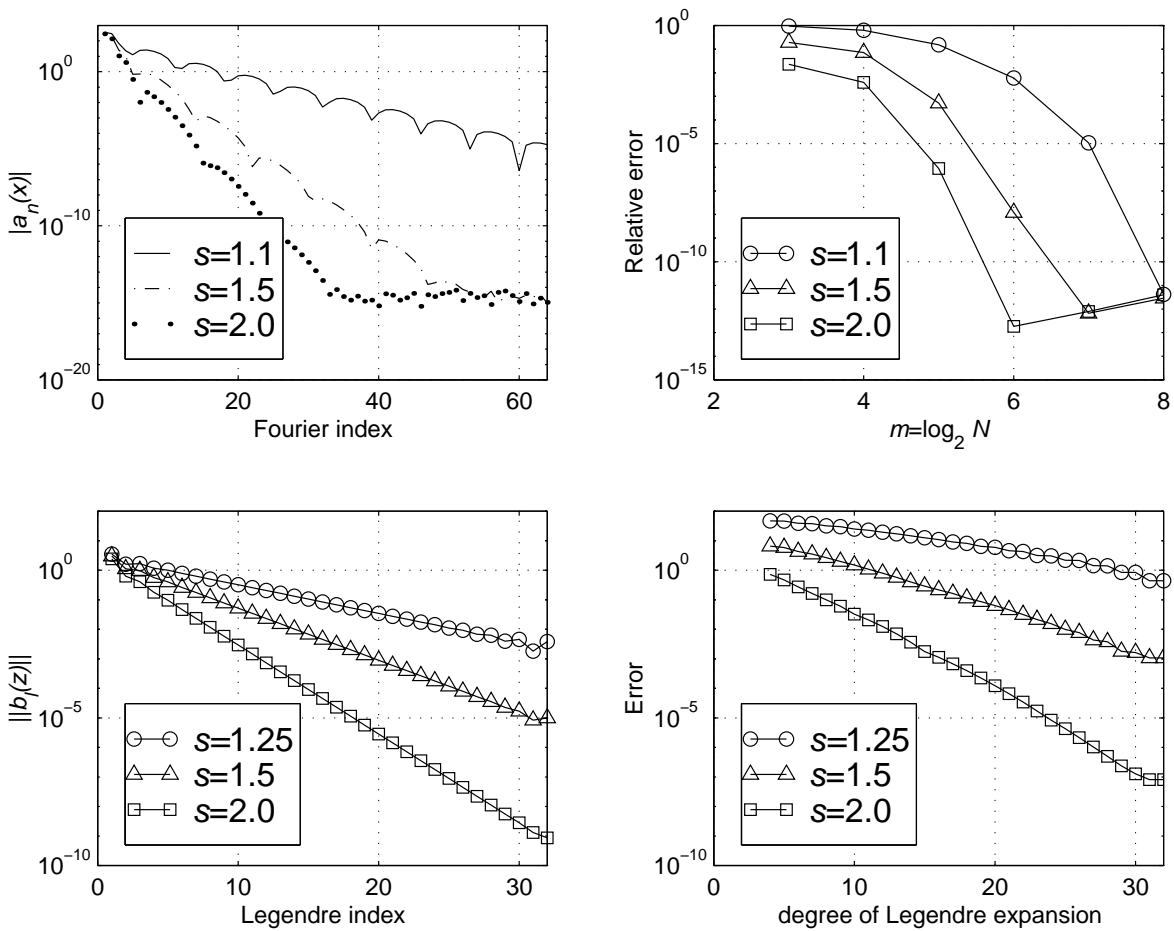}
\caption{The Boyer-Lindquist scalar field $\Phi(z,\theta)$ possesses a 
Fourier-Legendre expansion in $w=1/z, z=x+is$, ($top$ two windows) and 
$\cos\theta$ ($bottom$ two windows), here shown for a symmetric configuration 
consisting of two black holes of mass $M_{1,2}=1$ positioned at the north 
and south pole $p=\pm 1$ of a spherical coordinate system ($r,\theta,\phi$).
The Fourier coefficients show exponential decay by analyticity in 
Im$(z)\ge s>s^*(\theta)=p\sin\theta$ shown with $\cos\theta=0.25$ and 
$s^*=0.9682$. This gives rise to spectral accuracy in the first and second
coordinate derivatives ($top$ right). The norm of the Legendre coefficients 
shows exponential decay when $s$ is larger than $p$ ($left$). This gives rise to
spectral accuracy in the first and second derivatives with respect to $\theta$
($bottom$ right).}
\label{FIG_cicleE}
\end{figure*}
We apply these observations to Boyer-Lindquist initial data for 
two black holes \cite{coo91,pri94}, described by the potential
\begin{eqnarray}
\Phi=1+\frac{M_1}{\sqrt{z^2+p^2-2pz\cos\theta}}
                +\frac{M_2}{\sqrt{z^2+q^2-2qz\cos\theta}}
\label{EQN_PHI}
\end{eqnarray}
for a pair of black holes with masses $M_{1,2}$ at positions $p$ and $q$ along
the polar axis of a spherical coordinate system $(r,\theta,\phi$). It will be
appreciated that the singularities $z=pe^{\pm i\theta}$ are essential in making 
$\Phi$ non-trivial: the alternative of an entire function which is analytic at 
infinity would be constant by Liouville's theorem.  For complex
$z=x+is$ with $s$ larger than both $p$ and $q$, $\Phi(z,\theta)$ is finite 
and pointwise analytic everywhere on $-\infty+is<z<\infty+is$. Following (\ref{EQN_GZ})
and (\ref{EQN_ZK}), we capture analyticity by the
Fourier-Legendre transform on the image $w=1/z$. This gives rise to an efficient 
and accurate functional representation in $(z,\theta)$ with spectral accuracy
in derivatives by spectral differentiation.
Fig. 1 shows the computational results on the alternative representations
\begin{eqnarray}
\Phi(z,\theta)=\Sigma_{n=1}^N a_n(\cos\theta) w^n = \Sigma_{l=0}^L b_l(z) P_l(\cos\theta),
\label{EQN_ALT}
\end{eqnarray}
where the series are finite in our numerical implementations.

\begin{figure*}
\includegraphics[scale=0.50]{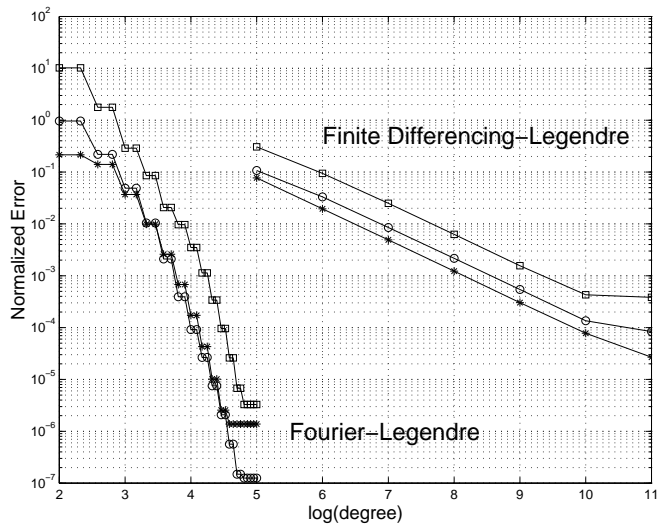}
\caption{Normalized errors in the first ({\em circles}) and second ({\em squares}) 
derivatives of the metric and the Ricci scalar tensor ({\em stars}) based on a 
Fourier-Legendre representation of Boyer-Lindquist initial data of Fig. 1 with 
$M/s=1/2$. The Fourier-Legendre results are computed with $N=32$ Fourier 
coefficients on $w=1/z$ varying degrees $L=4,\cdots,32$ of Legendre polynomials. 
The results are compared with a mized finite differencing-Legendre method with 
varying degree $N=2^m$, $m=5,\cdots11$ of the number of in the $z-$coordinate and 
fixed expansion in $L=16$ Legendre polynomials. This comparison highlights the
the accuracy and efficiency of the Fourier-spectral method, even at low resolutions.
The results are practically unchanged for other configurations of Boyer-Lindquist
data, and the errors decrease for larger $s$.}
\label{FIG_FL}
\end{figure*}
The application to black holes spacetimes in the complex plane can be
considered in the line-element 
\begin{eqnarray}
ds^2  = -N^2dt^2+h_{ij}dx^idx^j, ~~N(z)=\frac{1-M/2z}{1+M/2z},
\label{EQN_A1}
\end{eqnarray}
for a three-metric $h_{ij}$. Working in the complex plane, we are at liberty to 
choose a lapse function $N$ and vanishing shift functions in accord with the
asymptotic Schwarzschild structure of spacetime at large distances. This applies
without loss of generality to any binary black hole spacetime with total 
mass-energy $M$ as measured at infinity, defined by the residue
\begin{eqnarray}
M=\mbox{{\sc Re}}\left\{
\frac{i}{\pi}\int_{-\infty+is}^{\infty+is} 
\left(h_{zz}^{1/4}-1\right)dz\right\}.
\label{EQN_RES}
\end{eqnarray}
For binary black hole coalescence, $M$ is the sum $M_1+M_2$ of the masses of
the two black hole, plus the energy in gravitational radiation and minus the
binding energy in the system. Thus, $M$ is a constant. In contrast, $M$
is a non-increasing function of time on any finite computational domain
with outgoing radiation boundary conditions.
For large $z$, the line-element (\ref{EQN_A1}) satisfies
\begin{eqnarray}
ds^2= -N^2dt^2+\left(1+\frac{M}{2z}\right)(dz^2 + d\Sigma^2)+\cdots
\label{EQN_L1}
\end{eqnarray}
up to order $\left({M}/{z}\right)^2$,
where $d\Sigma^2=z^2d\theta^2 + z^2 \sin^2\theta d\phi^2$ denotes the line-element 
of the coordinate sphere of radius $z$. Accordingly, the three-metric $h_{ij}$
is decomposed into diagonal and off-diagonal elements given by the matrix 
factorization
\begin{eqnarray}
h_{ij}=\gamma\left(\begin{array}{cccc}
A_1^4 & B_{1} & B_{2}\\
B_{1} & A_2^4 & B_{3}\\
B_{2} & B_{3} & A_3^4
\end{array}\right)\gamma,
\label{EQN_L2}
\end{eqnarray}
where $\gamma$ is the diagonal matrix
\begin{eqnarray}
\gamma_{11}=1,
\gamma_{22}=z,
\gamma_{33}=z\sin\theta.
\end{eqnarray}
In accord with the spectral representation of (\ref{EQN_GZ}) and the potential
(\ref{EQN_PHI}), the coefficients $(A_i,B_i)$ are expanded in spherical 
harmonics on $z=x+is$,
\begin{eqnarray}
A_{i}=M\Sigma_{|m|\le l} ~a_{i}^{ml}(t)w^{l+1}Y_{ml}(\theta,\phi),
\label{EQN_L3a}
\end{eqnarray} 
\begin{eqnarray}
B_{i}=M\Sigma_{|m|\le l} ~b_{i}^{ml}(t)w^{l+1}Y_{ml}(\theta,\phi).
\label{EQN_L3b}
\end{eqnarray}
Together with the Fourier expansion on $w=M/z$, (\ref{EQN_L3a}-\ref{EQN_L3b})
defines a spectral representation in all three coordinates. Fig. (\ref{FIG_FL})
shows the results in case of the axisymmetric Boyer-Lindquist initial data,
including the computational error behavior in the scalar Ricci tensor. Here,
we use the general expansions (\ref{EQN_ALT}) to probe independently the
dependencies on $N$ and $L$ (to go beyond $N=L+1$ in any truncated
series of (\ref{EQN_L3a}-{\ref{EQN_L3b})).

\begin{figure*}
\includegraphics[scale=0.5]{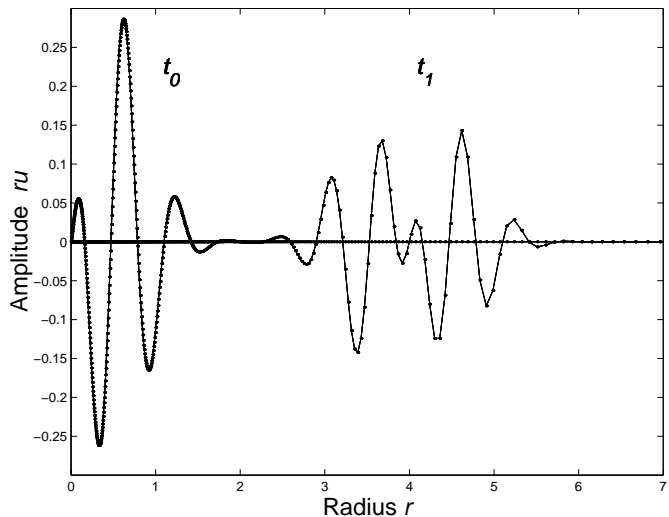}
\caption{
Signal recovery of a wave propagating to quiescent infinity is shown for a 
quantity $u(t,z)$ on $z=x+is$, here satisfying the scalar wave equation
$u_{tt}=r^{-2}(r^2u_{r})_{r}$ with time-symmetric initial data at $t=0$
$u(0,x)=e^{-2x^2}\cos(10x)$ and $u_t(0,x)=0$. The simulations use the
Fourier representation on the complex radius under $w=M/z$ and the
wave-equation $u_{tt}=w^4 u_{ww}$ using $N=1024$ points. Plotted is
the quantity $ru$ on $r\ge0$ for $t_0=0$ and $t_1=4$. The numerical results 
({\em dots}) accurately track the exact results ({\em continuous line}, by 
the method of characteristics). The errors are essentially independent of $s$, 
i.e., 4.5e-4 for $s=0.5$ and 4.7e-4 for $s=1.0$, as implied by analyticity
in $s>0$.}
\label{fig_w}
\end{figure*}
Time-evolution of $h_{ij}$ on $z=x+is$ gives rise to burst of radiation which
propagates towards quiescent infinity, where the observer's signal is defined 
on non-negative radii $z\ge0$. This signal can be calculated by projection according 
to Cauchy's integral formula,
\begin{eqnarray}
u(r,\theta,\phi,t)=\mbox{{\sc Re}}\left\{
\frac{i}{\pi}\int_{-\infty+is}^{\infty+is} \frac{u(z,\theta,\phi,t)}{z-r}dz
\right\},
\label{EQN_CIF}
\end{eqnarray}
where $u$ denotes a relevant metric component or tensor quantity.
Time-evolution on $z=x+is$ is stable when constant translation $s$, since this
preserves reality of the dispersion relation. This can be illustrated by the
linear wave-equation for a scalar field $u(t,r)$ in spherical coordinates 
$(t,r,\theta,\phi)$ with time-symmetric initial data, given by
\begin{eqnarray}
K: \left\{
\begin{array}{rll}
  u_{tt}(t,r) & =r^{-2}[r^2 u_{r}(t,r)]_r \\
  u(0,r)& = u_0(r),~~u_t(0,r)=0.
\end{array}
\right.
\label{EQN_K}
\end{eqnarray}
By analytic continuation, the equivalent initial value problem $K^\prime$ with 
complex radial coordinate on $-\infty +is < z < \infty+is$ is
\begin{eqnarray}
K^\prime: \left\{
\begin{array}{rl}
  u_{tt}(t,z) & = w^4 u_{ww} ~~(w=1/z)\\
  u(0,z) & =  u_0(z),~~u_t(0,z)=0.
\end{array}
\right.
\label{EQN_KP}
\end{eqnarray}
(In general, the analytic continuation of the initial
data is the extension of data on $(r,\theta,\phi)$ and $(r,\pi -\theta, \phi+\pi)$). 
We can implement $K^{\prime}$ numerically using finite-differencing
and leapfrog time-stepping. Fig. (\ref{fig_w}) shows a representative numerical result,
which is verified to satisfy independence of $s$.
The scalar wave-equation with time-symmetric initial data shows the appearance 
of both left and right movers, the first propagating towards $z=-\infty+is$ and
the second propagating towards $z=\infty+is$. Only the right
movers are projected onto the observer's non-negative radii $z\ge0$. The 
left movers are ultimately dissipated ``unseen" in their propagation to $-\infty+is$, 
while maintaining a contribution to the total mass-energy $M$ in (\ref{EQN_RES}). 

To conclude, the complex plane offers a unique opportunity for circumventing spacetime 
singularities and enabling spectral representations in all three dimensions, consisting
of spherical harmonics in the angular coordinates and Fourier series in the complex 
radial coordinate. This approach creates the most efficient representation of the metric,
promising a reduction in computational effort by orders of magnitude compared to standard
finite differencing approaches. It will be of interest to take advantage of this approach 
in time-dependent calculations described by the fully nonlinear equations for general 
relativity in any of its hyperbolic forms (e.g., \cite{van96}).

{\bf Acknowledgment.}
The author thanks the referee for constructive comments.
Part of this research is supported by the LIGO Observatories, constructed by 
Caltech and MIT with funding from NSF under cooperative agreement PHY 9210038.
The LIGO Laboratory operates under cooperative agreement
PHY-0107417.

\end{document}